

\magnification=\magstep1 
\font\bigbfont=cmbx10 scaled\magstep1
\font\bigifont=cmti10 scaled\magstep1
\font\bigrfont=cmr10 scaled\magstep1
\vsize = 23.5 truecm
\hsize = 15.5 truecm
\hoffset = .2truein
\baselineskip = 14 truept
\nopagenumbers
\input epsf.tex
%
\topinsert
\vskip 3.2 truecm
\endinsert
\centerline{\bigbfont Far From Equilibrium Dynamics of the Bose Gas}
\vskip 3 truept
\vskip 20 truept
\centerline{\bigifont {Kedar Damle}\ $^1$, {Satya N. Majumdar}\ $^2$, and 
{Subir Sachdev}\ $^1$} \vskip 8 truept
\centerline{\bigrfont 1.  Department of Physics}
\vskip 2 truept
\centerline{\bigrfont Yale University, New Haven, CT 06520-8120, USA}
\vskip 14 truept
\centerline{\bigrfont 2. Theoretical Physics Group}
\centerline{\bigrfont Tata Institute of Fundamental Research}
\centerline{\bigrfont Homi Bhabha Road, Mumbai-400005, India}
\vskip 1.8 truecm

\centerline{\bf 1.  Introduction}
\vskip 12 truept

The recent successful observations of Bose condensation in neutral,
trapped atomic gases [1] and excitons in $Cu_2O$ [2] have taken
the experiments into a heretofore inaccessible regime, as pointed
out by several speakers at this conference. These also have led theorists
to ask and study questions regarding the Bose gas which, without these 
recent success stories, would have been merely academic excercises.
One such exciting field is to understand the time-dependent nonequlibrium
phenomena in the Bose gas. In this paper, we study theoretically one
particular non-equilibrium question which we hope would be possible to
investigate experimentally in the near future. The question is the 
following: Following a rapid quench of the Bose gas from a high temperature
disordered state to a low temperature ordered state, how does the condensate
density grow from its initial value zero to its equilibrium value 
corresponding to the final temperature? A few recent papers [3,4] addressing
this question have focused on the early-time (of the order of a few collision
times) nonuniversal dynamics. However, as also noted in Ref. [5], the
interesting experimental questions are instead associated with the late time
dynamics that involves the coarsening of the Bose condensate order parameter
field. Our analytical and numerical studies of this late time dynamics of 
the Bose gas shows, in addition to answering the question on the growth 
of condensate density, that this dynamics represents a new universality
class of phase ordering kinetics.

Before elaborating on the nonequilibrium question, it would be useful
to recapitulate briefly the equilibrium behaviour of the Bose gas. A
dilute Bose gas (with repulsive interactions) at equilibrium undergoes
a phase transitions from a high temperature ``normal" state to a low
temperature ``superfluid" state at a nonzero critical temperature $T_c$
in space dimensions $d>2$. This low temperature superfluid state has
true long range order, i.e., the correlation function of the complex Bose 
order parameter field $\psi (r)$ approaches a nonzero constant,
$\langle \psi^{*}(0)\psi(r)\rangle \to |\langle \psi\rangle|^2$ as $r\to 
\infty$. On the other hand, in $d=2$, the Bose gas undergoes
a Kosterlitz-Thouless transition at $T=T_{KT}$. For $T< T_{KT}$, the system
has quasi long range order, i.e., the correlation function decays to zero
as $r\to \infty$ but in a slow power law fashion, $\langle \psi^{*}
(0)\psi (r)\rangle \sim r^{-\eta(T)}$ where the exponent $\eta (T)$
depends continuously on temperature ranging from $\eta(0)=0$ to 
$\eta(T_{KT})=1/4$. Associated with this phase transition what also 
happens in the low temperature ordered phase
is Bose condensation, namely, the average number of particles in the $k=0$
mode becomes macroscopic, i.e., $\langle n(k=0)\rangle =\langle 
|\psi( k=0)|^2\rangle \sim L^d$ for $d>2$ and $\sim L^{2-\eta(T)}$ for
$d=2$. It has been argued [6] that the statics of the Bose gas is in the
same universality class as that of the classical $XY$ model of ferromagnets.

We now come to the non-equilibrium question. Suppose that we prepare our
Bose gas at equilibrium at a very high initial temperature $T_i$ where the 
condensate density is zero and then rapidy quench the system to a 
temperature $T_f$ below the transition temperature. We start the clock
immediately after the quench and monitor the condensate density ${\rho}_0 
(t)=\langle n(k=0,t)\rangle/L^{d'}$ (where $d'=d$ for $d>2$ and
$d'=2-\eta(T)$ for $d=2$) as a function of time. As time progresses, this
density is expected to rise from its initial value $0$ and then eventually
saturate to its equilibrium value (a nonzero number of $\sim O(1)$). We
would like to know the precise form of this temporal evolution. We show 
below that 
this time evolution of the condensate density can be described in a natural
and precise language by using the phenomenology of phase ordering kinetics
developed recently for dissipative classical spin systems, as reviewed by
Bray [7]. As a byproduct of this study, we find that while the statics
of the Bose gas and classical $XY$ model are in the same universality
class, the non-equilibrium dynamics of these two systems belong to
different universality classes.

In Section-2, we briefly review the scaling hypotheses of the phase 
ordering theory and discuss the scaling predictions drawn from this 
general theory for the Bose gas. In Section-3, we introduce a solvable toy
model of coarsening of Bose gas in $d=2$ and present its exact solutions
to illustrate the importance of nondissipative Poisson bracket terms in the
equations of motion. Section-4 contains a discussion on the 
Gross-Pitaevski (GP)
equation that describes the evolution of the Bose gas in a ``microcanonical"
ensemble. Section-5 contains the numerical results on the $GP$ equation
in $d=2$ and $3$ that verifies the scaling predictions of Section-2. 
Finally, we summarize and conclude in Section-6. A shorter version of 
this work has appeared elsewhere [8].  

\vskip 28 truept

\centerline{\bf 2. Scaling Predictions From Phase Ordering Theory}
\vskip 12 truept

In the theory of phase ordering kinetics [7], one considers the evolution 
of a classical spin system (such as Ising model) after a rapid quench 
from a high temperature disordered phase to a low temperature ordered phase.
The dynamics is assumed to be overdamped and purely relaxational in which
each spin simply moves along the steepest downhill direction in the 
instanteneous energy landscape. Locally ordered regions will appear 
immediately
after the quench, but the orientation of the spins in each region will be
different. The coarsening process is then one of allignment of neighbouring
regions, usually controlled by the motion and annihilation of topological
defects (domain walls for Ising spins, vortices for $XY$ spins, etc.). 
As time progresses, these domains grow in size as the system tries to 
achieve local ordering on larger and larger scales. A key step in the
theory is the introduction of a single length scale $l(t)$, a monotonically
increasing function of the time $t$, which is about the size of a typical
ordered domain at time $t$. At late times when $l(t)$ is larger than the 
microscopic length scales such as the range of interactions or the lattice
spacing, it is believed that the late stage morphology of the system is 
completely characterized by $l(t)$, and is independent of microscopic 
details or initial conditions (as long as the initial condition is short 
ranged), i.e., it is universal. At late times, $l(t)$ typically grows like
a power law, $l(t)\sim t^{1/z}$ where the exponent $z$ depends upon the
various conservation laws satisfied by the dynamics. For nonconserved
Ising or $XY$ spin dynamics, it is well established that $z=2$ [7].

The morphology of growing domains is characterized by various time dependent 
correlation functions which exhibit universal scaling behaviour. For the
purpose of comparision with the Bose gas, let us illustrate these scaling
predictions for the $2$ component classical $XY$ model. Let $\psi(r,t)$
denote the order parameter field for the $XY$ model. Then in an infinite 
system, the scaling 
hypothesis of phase ordering kinetics predicts a scaling form for the
equal time correlation function, $G(r,t)=\langle \psi(0, t)\psi
(r,t)\rangle \sim r^{-\eta}g[rt^{-1/z}]$ where $\eta =0$ for $d>2$ and in
$d=2$, $\eta=\eta(T)$, the usual temperature dependent exponent associated
with the final equilibrium state. This means that the structure factor, the
Fourier transform of $G(r,t)$, will scale as $S(k,t)\sim t^{(d-\eta)/z} 
{\tilde g}(kt^{1/z})$. In particular, the $k=0$ mode will grow as
$S(0,t)\sim t^{(d-\eta)/z}$ in an infinite system.

However, in a finite system of linear size $L$, the system will stop
coarsening and attain the equilibrium ordered state when $l(t)\sim L$.
In that case, these scaling behaviour of the inifinite system would be 
modified by finite size scaling (FSS). For example, it would then predict
a FSS form for the equal time correlation function, $G(r,t)\sim L^{-\eta}
{\Phi}_{G}[r/L, t/L^z]$ and hence $S(0,t)\sim L^{d-\eta}\Phi [t/L^z]$. This
last scaling function $\Phi$ goes to a constant for $t>>L^z$ and the system
attains equilibrium after $t\sim L^z$. The value of $z$ is known to be $2$ 
for the classical $XY$ model [7] apart from the logarithmic corrections 
in $d=2$ [15,16].

Now consider the Bose gas. The order parameter in this case is the boson 
annihilatin field $\psi (r,t)$ which is complex; the phase of the
expectation value of $\psi$ is aligned across the system in the 
equilibrium Bose-condensed state. A key point is that after relatively 
few atomic collisions, once the domain size $l(t)$ is large enough (e.g., 
larger than the de Broglie wavelength), it is permissible [4] to treat 
$\psi(r,t)$ as a {\sl classical} field which obeys Hamilton-Jacobi 
equations of motion ( for a related discussion on the emergence of 
classical dynamics in the equilibrium properties of an antiferromagnet, see
Ref. [9]). It must be kept in mind that it is only the equations of 
motion for the collective order parameter which are classical-the very 
existence of the complex order parameter is due entirely to quantum 
mechanics, and the fact that there is a wavefunction for the condensate..

If we believe that the late time ordering dynamics of the Bose field
can be described by the scaling phenomenology of the phase ordering theory
described above (even though the equation of motion for the Bose field
is different from that of the classical spins), then one obtains the
prediction that the $k=0$ mode of the structure function, which in
the Bose gas case is just the number of particles in the $k=0$ mode, will
scale as $S(0,t)\sim L^{d-\eta}\Phi[t/L^z]$. Then once we know the 
exponent $z$ and the scaling function $\Phi$, it will give us the
temporal evolution of the condensate density, the question we started
out with. 

Of course the value 
of the exponent $z$ and the scaling function $\Phi$ may be different
from the classical $XY$ case as they depend on the conservation laws 
satisfied by the equations of motion of the order parameter.
Indeed one of the central results of this paper is to establish that
while this scaling prediction holds for the Bose gas, the exponent
$z$ in the Bose gas (evolving via the GP equation as discussed later) is 
different from that of the $XY$ model. This
is due to the important property of the equations of motion for $\psi$,
discussed in Section-4, that they are not simply relaxational. Instead,
they contain nondissipative, kinematical ``streaming" or ``Poisson bracket" 
terms [10]. One such term is responsible for the Josephson precession of
the phase of $\psi$ at a rate determined by the local chemical potential.  
One of our main objectives is to understand the consequences of such 
terms on the phase ordering theories. We will argue that the Josephson term
constitutes a relevant perturbation on the dynamics and that the 
resulting coarsening process belongs to a new universality class.

In Section-5, we give numerical evidence supporting the scaling of
$S(0,t)$ as predicted above and also determine the exponent $z$ 
numerically. But before doing that, in the next Section, we consider a 
solvable toy model of coarsening which will prove that indeed the 
``streaming terms" in the equation of motion are relevant perturbations. 
This will prepare us to expect that the value of $z$ in the Bose gas 
dynamics might be different from the classical $z=2$ of the $XY$ model
and also give us important physical insights as to why they are different.
\vskip 28 truept

\centerline{\bf 3. A Solvable Toy Model of Coarsening}
\vskip 12 truept 

In this Section we consider a simple toy model of coarsening that 
illustrates the possible consequences of the Josephson term in a simple 
setting. Consider the Bose gas in $d=2$. As mentioned earlier, for 
$T<T_{KT}$, the Bose gas is superfluid. Now consider the phase ordering
process in which the Bose gas is rapidly moved at time $t=0$ from contact 
with a reservoir at an initial $T=T_i$, to a reservoir with a final 
$T=T_f$, such that $T_f < T_i < T_{KT}$. A similar quench was considered 
in Ref. [12] for the purely dissipative $XY$ model. As time progresses, the
system will approach the equilibrium configurations corresponding to
$T=T_f$ starting from the initial configurations corresponding to $T_i$; this
ordering will proceed simply via the spin wave dynamics. This is in contrast
to the ordering via annihilation of vortex-antivortex pairs as in the case
of quench from a temperature above $T_{KT}$. Since there are no vortices
in the initial configurations, they won't be generated because the system
will only reduce its energy. 

Indeed in the long time limit, all vortices
and fluctuations in the amplitude of $\psi$ can be neglected, and we may
parametrize $\psi=\exp (i\phi)$. The free energy density in the purely
dissipative $XY$ model is now determined simply by the gradients of the 
phase $\sim (\nabla \phi)^2$. In the case of the Bose gas, it is also 
necessary to take the conserved number density into account. Let $m$ be 
proportional to the deviation of the particle density from its mean 
value; then the free energy density we shall work with is
$$ 
{\cal F} ={1\over {2}}\int d^2r [(\nabla \phi)^2 + m^2]. ~~\eqno(1)
$$

We have rescaled spatial coordinates and $m$ to obtain convenient 
coefficients in $\cal F$. Note that the fields $m$ and $\phi$ are
not independent but are related via the Poisson bracket
$$
\{ m(r), \phi(r')\}=g_0 \delta(r-r'), ~~\eqno(2)
$$
where $g_0$ is a constant. The origin of the Josephson precession term,
whose effects on dynamics we wish to study, lies in this Poisson bracket.
The method reviewed in Ref. [10] now leads to the {\sl linear} equations
of motion [6]
$$
{{\partial \phi}\over {\partial t}}={\Gamma}_0{\nabla}^2\phi +g_0m +\theta,
~~\quad\quad {{\partial m}\over {\partial t}}={\lambda}_0{\nabla}^2m 
+g_0{\nabla}^2\phi +\zeta,~~\eqno(3)
$$
where the coefficients $\Gamma_0, \lambda_0>0$ represent the dissipation 
arising from coupling of the system to the reservoir. The effects of the 
reservoir are also contained in the Gaussian thermal noise sources 
$\theta$ and $\zeta$ with zero mean and (for $t>0$) correlations 
appropriate to $T=T_f$: $\langle 
\theta(r,t)\theta(r',t')\rangle=2{\Gamma_0}T_f\delta(r-r')\delta(t-t')$, 
$\langle 
\zeta(r,t)\zeta(r',t')\rangle=-2\lambda_0T_f
{\nabla}^2\delta(r-r')\delta(t-t')$,
and $\langle \zeta(r,t)\theta(r',t')\rangle=0$ ($k_B=1$).
Equations (3) are linear, can be easily integrated and all correlations
can be computed exactly.

Let us first recall the structure of the solutions expected from naive 
scaling [7] for $d=2$. One expects a single length scale growing as 
$l(t)\sim t^{1/z}$. Also the morphology of the evolving patterns 
are characterized by two types of correlation functions: (i)The equal
time correlator $G(r,t)=\langle {\psi}^{*}(r,t)\psi(0,t)\rangle$ is
expected to scale as $G(r,t)\sim r^{-\eta_f}g(r/t^{1/z})$ where $g$ is a 
universal scaling function and $\eta_f$ is the equilibrium exponent of 
the quasi long range order at $T=T_f$ as mentioned in Section-2. (ii)The
unequal-time correlation function 
$C(r,t)=\langle{\psi}^{*}(r,t)\psi(0,0)\rangle$ for which we expect for 
large $r$ and $t$, $C(r,t)\sim t^{-{\lambda}/z}f(r/t^{1/z})$ where $f$ is 
a universal scaling function, and $\lambda$ is the autocorrelation exponent. 

It turns out that our model ${\cal F}$ does not completely obey the simple 
scaling hypotheses as stated above. This becomes clear upon considering 
the two-time correlator $C$ whose explicit exact solution turns out to 
depend upon ${\sl two}$ time-dependent length scales $l_1(t)\sim (at)^{1/2}$
and $l_2(t)\sim g_0t$ (with $a=(\Gamma_0+\lambda_0)/2$). It actually 
obeys the scaling form $C(r,t)\sim t^{-(3\eta_i+\eta_f)/4}{\tilde 
f}[r/{(at)^{1/2}}, r/(g_0t)]$ (where $\eta_i=T_i/{2\pi}$). The dependence 
of these scales on $g_0$ suggests that $g_0$ is a relevant perturbation 
with renormalization group eigenvalue $1$, in the language of Ref. [7]. 
The scaling function ${\tilde f}$ is found to be
$$
{\tilde f}(x_1,x_2)=\exp\bigl[-{\eta_i}\int_0^{\infty}{dy\over 
{y}}\{1-J_0(y)\}\cos (y/x_2)e^{-y^2/{x_1}^2}\bigr]. ~~\eqno(4)
$$
For $r\sim l_1(t)$, using ${\tilde f}(x_1, x_2\to 0)=1$, we find that the
autocorrelation $C(0,t)\sim t^{-(3\eta_i+\eta_f)/4}$ in contrast to the 
result in the model of Ref. [12] $C(0,t)\sim t^{-(\eta_i+\eta_f)/4}$. On 
the contrary, one could insist on a scaling picture using only the single 
larger length scale $r\sim l_2(t)$, and would then need ${\tilde 
f}(x_1\to \infty, x_2)$ which equals $[1+{\sqrt {1-{x_2}^2}}]^{-\eta_i}$ for
$x_2<1$ and equals ${x_2}^{-\eta_i}$ for $x_2>>1$. It can also be checked 
that one recovers the initial equal time equilibrium result for $C(r,t)$ 
when $r\to \infty$ with $t$ large but fixed. We also note that the 
relevance of $g_0$ is evident in the autocorrelations of $m$. We find 
$\langle m(0,t)m(0,0)\rangle \sim {1\over {t}}f_1(g_0{\sqrt {t/a}})$ where
$$
f_1(\tau)=4{\pi}^2\eta_i\bigl[1-\int_0^{\infty}{\sin 
y}e^{-y^2/{2{\tau}^2}}dy\bigr]; ~~\eqno(5)
$$
clearly, for $g_0=0$, this autocorrelator decays as $1/t$ for large $t$, 
while for nonzero $g_0$ it decays faster as $t^{-2}$. Finally, results on 
the equal-time $\psi$ correlator $G$ are as follows. It has a crossover 
time $t_1\sim {\tilde a}/{{g_0}^2}$ with ${\tilde 
a}=|\Gamma_0-\lambda_0|$; this time is similar to the crossover time in 
$\langle m(0,t)m(0,0)\rangle$, except that ${\tilde a}$ has replaced $a$. 
Both for $t<<t_1$ and for $t>>t_1$, $G$ obeys a scaling form similar to 
that obtained in Ref. [12] (which has $g_0=0$): $G(r,t)\sim 
r^{-\eta_f}g(r/{(\gamma t)^{1/2}})$ where $g$ is the scaling function 
described in [12]; however, the rate $\gamma=\Gamma_0 $ for $t<<t_1$ and
$\gamma=a$ for $t>>t_1$.

While this phase only model ${\cal F}$ is not relevant for studying 
quenches from above the transition temperature (since it neglects the 
nonlinear terms and hence the vortices which are the elementary defects
for the quench from high temperatures), the exact solution of this linear 
model is quite instructive. It clearly emphasizes the importance of the 
nondissipative Josephson coupling term. In fact as seen above, the 
presence of this term ($g_0\ne 0$) changes the universality class of the 
dynamics. Thus it is reasonable to expect that even for quenches from 
above the transition temperature, this term would play an important role.
In fact this is what we demonstrate in the next section by studying the
full nonlinear equation that describes the time evolution of the order
parameter.
\vskip 28 truept

\centerline {\bf 4. Coarsening of Bose Gas: A Deterministic 
Microcanonical Approach}
\vskip 12 truept

Consider the quench of the Bose gas from above the transition temperature.
In this case, the initial configuration, being a typical high 
temperature configuration, contains several different 
types of topological defects (excitations). 
As the system dissipates energy with time,  
most of these defects will disappear after a short transient time leaving
behind only the elementary excitations. In $d=2$, these elementary 
excitations are
point vortices and in $d=3$ they are vortex lines. As time progresses, these
elementary defects move around and annihilate each other upon meeting (and
thereby release energy) and the system becomes more and more ordered. To 
study this coarsening process that proceeds via the annealing of defects 
it is necessary to study the evolution of both the phase and amplitude of 
$\psi$. 

This growth of long range order in the system can be studied in two ways. 
In one case one considers the deterministic
time evolution of an {\sl isolated} Bose 
gas, not in contact with a heat bath. What we find in our study is that
{\sl though the dynamics in this case is nondissipative, the system still 
exhibits an irreversible approach to the equilibrium}. In the other case, 
the Bose gas is in contact with a heat bath and its evolution
equations are therefore necessarily stochastic. These are the analogues of 
{\sl microcanonical} and {\sl canonical} ensembles in equilibrium 
statistical mechanics. Most previous studies on coarsening have been done 
in the stochastic 
``canonical" ensemble. While it may be reasonable to expect that both 
descriptions may lead to same results for the universal scaling properties, 
a word of caution, however, is warranted since this equivalence is well 
established only for equal time properties of 
equilibrium systems. Equilibrium and nonequilibrium dynamics may be 
more subtle; indeed, in a recent study [11] 
of unequal time dynamics of the quantum
Ising chain in a transverse field it was found that the underlying
deterministic quantum dynamics did not map onto any known classical
stochastic model.

In this paper, we use only the determininstic 
``microcanonical" approach and do not 
make any statement about the ``canonical" results. To 
the best of our knowledge, this deterministic
``microcanonical" approach has never been 
used before to study coarsening in any system. The use of this approach
is not just cosmetic, 
in fact it has some advantages over the ``canonical" 
approach, atleast for the Bose gas. As we will see below, the dynamics in 
the ``microcanonical" 
approach is completely specified by the Hamiltonian of the system with no 
additional phenomenological parameters. The ``canonical" dynamics, on the 
other hand, needs several phenomenological constants as input parameters.
Therefore, the ``microcanonical" dynamics is much easier to implement
numerically and one does not need to do a ``time consuming" search in a
rather big parameter space as in the ``canonical" case.

For the isolated Bose gas (``microcanonical" ensemble), an excellent 
approximation for the total energy of an order parameter configuration 
$\psi (r,t)$ is ${\cal H}=\int d^dr [ |\nabla \psi|^2 +{u\over 
{2}}|\psi|^4]$, where the length scales have been rescaled to make the 
coefficient of the gradient term unity, and $u>0$ is the two-particle $T$ 
matrix at low momentum, representing the strength of the repulsive onsite
interaction. The standard Hamilton-Jacobi equation of motion for $\psi$ 
follows using the Poisson bracket $\{\psi, {\psi}^{*}\}=i$
$$
i{{\partial \psi}\over {\partial t}}=[-{\nabla}^2 +u|{\psi}|^2]\psi, 
~~\eqno(6)
$$
and is well known [13] Gross-Pitaevski (GP) or nonlinear Schrodinger 
equation. We can also add a quadratic $|\psi|^2$ term to ${\cal H}$, and 
it leads to a term linear in $\psi$ in the GP equation; however this 
linear term can be eliminated by an innocuous global phase change in 
$\psi$. The GP equation conserves the total number of particles $\int 
d^dr |\psi|^2$, the total momentum, and ${\cal H}$, and hence there is
no global dissipation of energy. Nevertheless, in the thermodynamic 
limit, the GP equation does display irreversible coarsening, as will be 
be abundantly clear from our numerical results to be described in the 
next Section. A random initial state with a negiligible number of 
particles in the zero momentum $(k)$ state (i.e., short range initial 
correlations), evolves eventually to a state with a condensate fraction 
equal to that expected at equilibrium in the microcanonical ensemble at 
the total energy of the initial state. Basically while the total energy 
of the system is conserved, there is nevertheless an energy flow from the
low momentum states to high momentum states thus effectively making the
system more and more ordered as time progresses.

In the ``canonical" approach on the other hand, it is permissible to add 
dissipative terms to the equation of motion of $\psi$. A simple 
additional damping term to the GP equation leads to a model expected to be 
in the same universality class of the so-called Model-A [10,7]; this 
model is, however, not acceptable: it violates the local conservation of 
the particle density, and, as discussed before Eq. (3), it is necessary 
[10,14] to introduce the density fluctuation field, $m(r,t)$; the value 
of $|\psi(r,t)|^2$ is then the contribution to the particle density from 
the low momentum states, while $m(r,t)$ represents the density fluctuation
from all states; the Poisson bracket in this case is $\{m(r), 
\psi(r')\}=ig_0\psi(r)\delta(r-r')$. This is model F in the language of 
Ref. [10].
(It is probably also necessary to introduce additional fields to account for other 
conserved quantities: a momentum density as in Model H or an energy 
density as in Model C of Ref. [10].)
Note that the strength of the crucial 
precession term in the dynamics is controlled by $g_0$ which is an 
adjustable phenomenological parameter (however, in the Hamiltonian 
dynamics of the microcanonical approach, there is no such freedom). 
Numerical study of coarsening using model $F$ could thus be complicated by 
crossover effects associated with the adjustable value of $g_0$ ($g_0=0$ 
corresponds to the purely dissipative model-A dynamics, which is clearly 
in a different universality class).

We therefore restrict our numerical study here to the ``microcanonical"
approach to coarsening using the GP equation. These results are the 
subject of the following Section.
\vskip 28 truept

\centerline {\bf 5. Numerical Results }
\vskip 12 truept

All of the numerical results obtained so far are consistent with the 
simplest naive scaling hypotheses described earlier, and do not require 
the introduction of two length scales, as was necessary in the linear 
model of Section-3 (though we have not yet obtained numerical results on 
unequal-time correlations, for which the linear model ${\cal F}$ clearly 
displayed two length scales). We will present results both in $d=2$ and 
$d=3$. The $d=2$ system allowed us to study larger sizes with better 
finite-size scaling properties.

We discretized Eq. (6) on a lattice, and integrated in time using a fast 
Fourier transform based algorithm which conserved energy and particle 
number to a high accuracy. We work in units where the lattice spacing is 
unity and choose the scale of the lattice field to make the number density 
unity also. We set $u$ to be approximately $0.25$ so that we are 
considering a dilute gas. We choose an ensemble of initial conditions with
a narrow distribution of energy, whose width goes to zero in the 
thermodynamic limit. We assign initial values to the Fourier components 
$\psi(k,0)$ as follows: $\psi(k,0)={\sqrt n_0(k)}\exp[i\phi(k)]$ where 
the $\phi(k)$'s are independent random variables chosen from a uniform 
distribution with range $[0,2\pi]$ and the function $n_0(k)$ is chosen to 
ensure that initial real-space correlations are short ranged 
(corresponding to a ``high-temperature" configuration) while still having 
low enough energy so that the equilibrium state corresponding to this 
energy is superfluid. Though the ensemble of initial conditions is not 
strictly the Gibbs distribution at any temperature, it is however 
expected that the precise details of the initial conditions do not matter 
for the late time universal properties as long as the initial 
correlations are short ranged. More specifically we chose
$$
n_0(k)= {c\over { [\epsilon (k)+{\mu}_1]}} {1\over {[ 1+\exp\{(\epsilon 
(k)-\mu_2)/T\}]}}, ~~\eqno(7)
$$
where $\epsilon (k)$ is the Fourier representation of the lattice version 
of the Laplacian and $c$ sets the overall scale of $n_0(k)$. Here one 
chooses the parameters $\mu_1$, $\mu_2$, and $T$ to achieve the 
appropriate trade-off between energy and correlation length. Note that 
this careful choice of initial conditions is needed as the GP equation 
does not have any explicit dissipation and the system evolves in the 
phase space on a constant\break
\topinsert
\epsfxsize=3.5in
\centerline{\epsffile{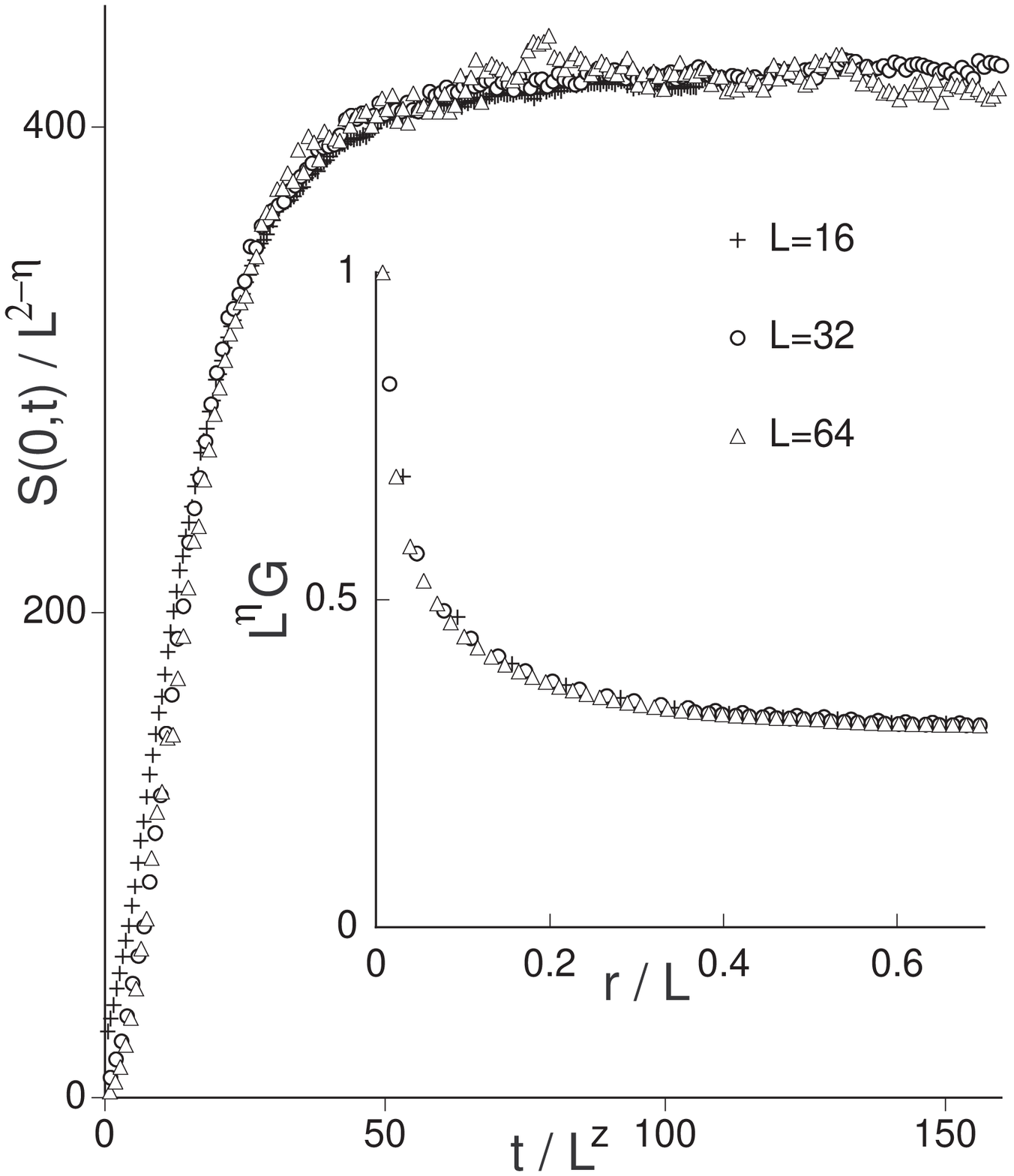}}
\noindent
{\bf Figure 1.} Numerical results from the simulation of GP equation in 
$d=2$. The number of particles in the zero momentum state is $S(0,t)$ and 
the figure shows its scaling properties as a function of the system size 
$L$ and time $t$. The inset shows the scaling of the equilibrium 
equal-time correlation function $G(r, t\to \infty)$. The best scaling 
collapse was obtained in both plots for $\eta \approx 0.27$ and $z\approx 
1.1$. The scale of all axes (except the values of $r/L$) are arbitrary. 
\endinsert
\vskip 28 truept
\noindent
energy surface. So, one has to choose this
constant energy surface
via tuning the initial conditions in such a way that ensures that there are
indeed some long-ranged configurations on this surface where the system can
finally go to. The point is that if there are such long-ranged 
configurations on the constant energy manifold then the dynamics of the 
system evolving via the GP equation takes the system automatically and
irreversibly (in the thermodynamic limit) to those long-ranged 
configurations. So the choice of this complicated initial condition just 
ensures that there are indeed such long-ranged configurations on the 
constant energy surface.

We tested the finite size scaling form mentioned in Section-2 for the equal 
time correlator: $G(r,t)=L^{-\eta}\Phi_G[r/L, t/L^z]$ where $\eta=0$ in
$d=3$ and in $d=2$, $\eta$ is the exponent associated with the final 
equilibrium state. We also computed the number of particles in the $k=0$
mode $S(0,t)$ expected to scale as $S(0,t)\sim L^{2-\eta}\Phi[t/L^z]$ as
mentioned also in Section-2.

Results for $d=2$ are shown in Fig. (1). We performed finite-size
scaling\break 
\topinsert
\epsfxsize=3.5in
\centerline{\epsffile{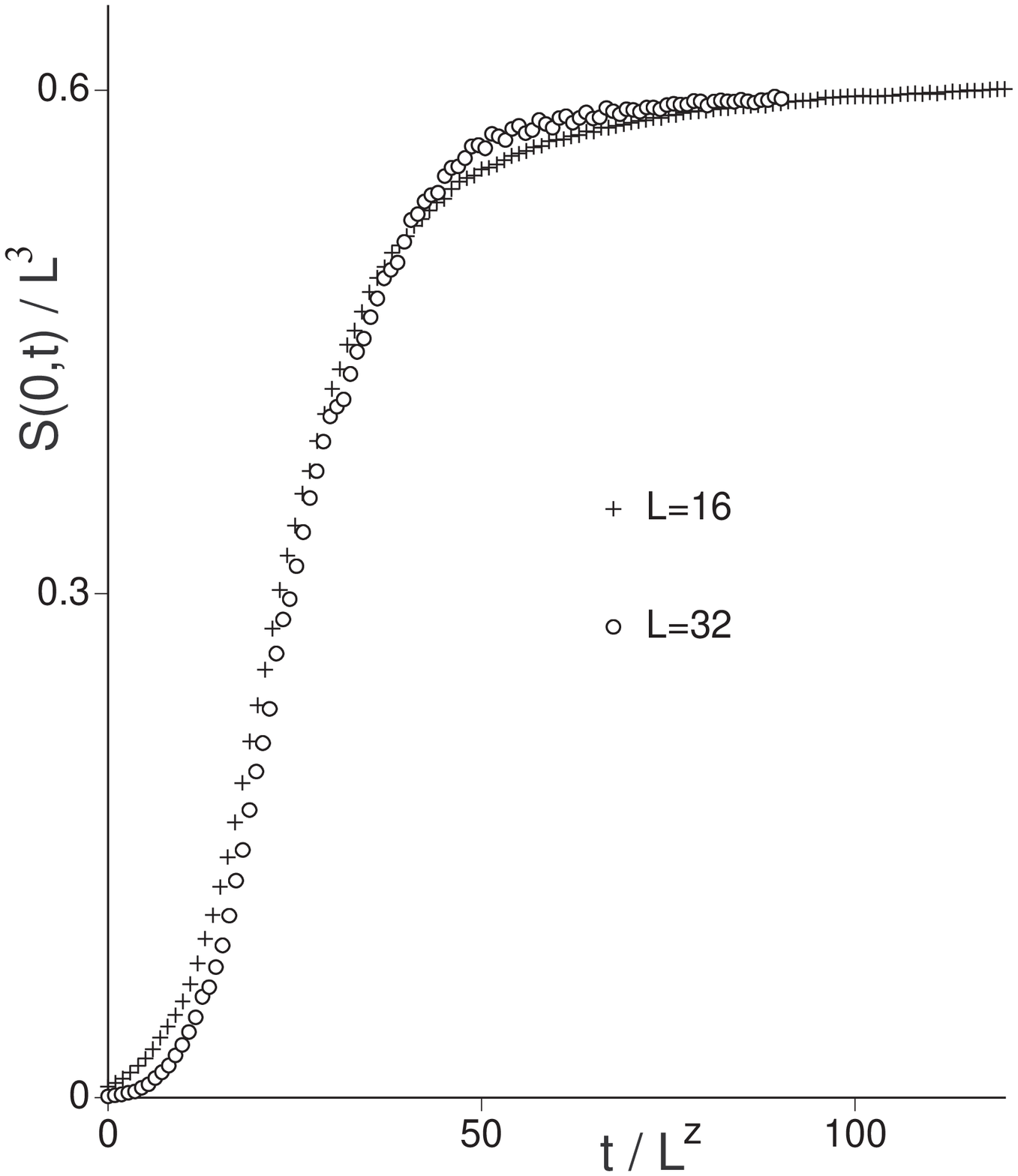}}
\noindent
{\bf Figure 2.} Numerical results for the GP equation in $d=3$. The 
notation is as in Fig. 1, with the exponent $z\approx 1.15$.
\endinsert
\vskip 28 truept
\noindent
analysis for $L=16$, $32$, and $64$ and found reasonable data collapse 
with $\eta\approx 0.27$ and $z\approx 1.1$. The value of $\eta$ indicates 
that we are at a nonzero temperature close to $T_{KT}$; strictly speaking 
we must have $\eta \le 1/4$, but the value of $\eta$ is relatively $T$ 
independent near $T_{KT}$, and the discrepancy is within our numerical 
errors. The value of $z$ is in sharp contrast to the 
$z=2$ (with logarithmic corrections)
result obtained by various groups [15,16] for 
the purely dissipative Model-A dynamics [10] (obtained from Model F by 
setting $g_0=0$ and ignoring the $m$ field) of classical $XY$ spins. 
Although we have determined the value of $z$ for a quench to a 
particular temperature $T_f=2\pi\eta\approx 1.695$ (in units of $k_B=1$), we 
expect that $z$ is same for all $0<T_f<T_{KT}$. 
Results for $d=3$ are shown in Fig. (2) for linear sizes $L=16$ and $32$. 
The data collapse is not as 
good as in $d=2$, but again we obtained a $z\approx 1.1$. Thus our 
numerical results, both in $d=2$ and $3$, are consistent with a value of 
$z=1$, which is also the result suggested by the exact calculation in the 
phase only model ${\cal F}$ of Section-3.

\vskip 28 truept

\centerline {\bf  6. Conclusion}

We close with some physical discussion on reasons for the difference 
between the deterministic
GP model, and quenches in the stochastic and purely dissipative Model A 
[15,16]. The dynamics in the GP model proceeds via the annihilation of 
nearby vortex-antivortex pairs (in $d=2$) as in Model A. However there is
an important difference between the two in details of the vortex motion. 
In Model A, oppositely charged vortices attract each other with a force 
that falls off as the inverse of their separation (apart from logarithmic 
corrections). The overdamped dynamics causes the vortices to then
move towards each other with a velocity proportional to
attractive force, and this implies $l(t)\sim 
t^{1/2}$. In the GP model, on the other hand, the situation is much more 
complex. In addition to vortices, the system also has a propagating 
``spin-wave" mode arising from the ``streaming" terms in the equation of 
motion. The finite velocity of this propagating mode gives rise to the
linear length scale $l(t)\sim t$. A 
pair of oppositely charged vortices, apart from
interacting with the spin-wave background, also has an attractive force
between them. However, now the underlying dynamics 
causes the pair to move with uniform 
velocity in a direction perpendicular to the line joining them
(the force leading to this motion is often called the Magnus force). 
These 
qualitative differences in the nature of the defect dynamics change the 
universality class of the coarsening process of the Bose evolving via
the GP equation.

In summary, we have presented evidence, both analytical and numerical, 
that the phase-ordering dynamics of an isolated Bose gas belongs to a new 
universality class. A particular conclusion of this work is that the 
condensate density of the Bose gas, following a sudden quench from the 
normal to the superfluid phase in dimensions $d\ge 2$, will grow at late 
times in a power law fashion as $t^{d/z}$ before saturating to its final 
equilibrium value. Our work, both analytical and numerical, provide
evidence that $z=1$ for the Bose gas evolving via the GP equation. Whether
this value of $z$ is same as that of the ``canonical" Model-F of Ref. [10]
remains an open question. In fact, recent numerical results on ``canonical",
stochastic
Model-F do indicate that the value of $z$ might be $2$ for that case [17].
(We speculate this may be because while Model-F has accounted for the
conserved number density, it has not accounted for the conserved
energy and momentum densities of the GP dynamics.) 
While the value of $z$ may be different for different dynamics, the scaling
prediction of the power law growth of the condensate density $\sim t^{d/z}$
at late times remains valid and needs to be tested experimentally. How this
simple scaling gets modified in presence of harmonic traps has been 
discussed in Ref. [18].

Given the latest advances in the experiments on Bose systems as we heard 
from various exciting talks in this conference, we may conclude with the 
hope that experimental verifications of our theoretical predictions 
summarized in the preceding paragraph may not be far off.

We thank D. Kleppner for stimulating our interest in this problem, and 
M.P.A. Fisher, A. Bhattacharya, A. Chakravarty and E. Cornell for 
useful discussions. This reserach was supported by NSF Grants 
DMR-96-23181 and DMR-91-20525.
\vskip 28 truept

\centerline{\bf References}

\vskip 12 truept

\item{[1]} 
M.H. Anderson et. al., Science {\bf 269}, 198 (1995); C.C. Bradley et. al.,
Phys.\ Rev.\ Lett. {\bf 75}, 1687 (1995).
\smallskip

\item{[2]}
D.W. Snoke, J.P. Wolfe, and A. Mysyrowicz, Phys.\ Rev.\ Lett. {\bf 64}, 
2543 (1990); Phys.\ Rev.\ B {\bf 41}, 11171 (1990); J.-L. Lin and J.P. 
Wolfe, Phys.\ Rev.\ Lett. {\bf 71}, 1223 (1993).
\smallskip

\item{[3]}
H.T.C. Stoof, Phys.\ Rev.\ A {\bf 45}, 8398 (1992); B.V. Svistunov, J.\ 
Moscow\ Phys.\ Soc. {\bf 1}, 373 (1991); D.V. Semikoz and I.I. Tkachev, 
Phys.\ Rev.\ Lett. {\bf 74}, 3093 (1995).
\smallskip

\item{[4]}
Yu.M. Kagan, B.V. Svistunov, and G.V. Shlyapnikov, Zh.\ Eksp.\ Teor.\ 
Fiz. {\bf 74}, 279 (1992) [Sov.\ Phys.\ JETP {\bf 75}, 387 (1992)].
\smallskip

\item{[5]}
Y. Kagan and B.V. Svistunov, Zh.\ Eksp.\ Teor.\ Fiz. {\bf 105}, 353 
(1994) [Sov.\ Phys.\ JETP {\bf 78}, 187 (1994)].
\smallskip

\item{[6]}
D.R. Nelson and D.S. Fisher, Phys.\ Rev.\ B {\bf 16} 4945 (1977).
\smallskip

\item{[7]}
A.J. Bray, Adv.\ Phys. {\bf 43}, 357 (1994).
\smallskip

\item{[8]}
K. Damle, S.N. Majumdar, and S. Sachdev, Phys.\ Rev.\ A {\bf 54}, 5037 
(1996).
\smallskip

\item{[9]}
S. Chakravarty, B.I. Halperin, and D.R. Nelson, Phys.\ Rev.\ B {\bf 39}, 
2344 (1989); A.V. Chubukov, S. Sachdev, and J. Ye, {\it ibid.} {\bf 49}, 
11919 (1994).
\smallskip

\item{[10]}
P.C. Hohenberg and B.I. Halperin, Rev.\ Mod.\ Phys. {\bf 49}, 435 (1977).
\smallskip

\item{[11]}
S. Sachdev and A.P. Young, Phys.\ Rev.\ Lett. {\bf 78}, 2220 (1997).
\smallskip

\item{[12]}
A.D. Rutenberg and A.J. Bray, Phys.\ Rev..\ E {\bf 51}, R1641 (1995).
\smallskip

\item{[13]}
E.P. Gross, Nuovo Cimento {\bf 20}, 454 (1961); L.P. Pitaevski, Zh.\ Eksp.\ 
Teor.\ Fiz. {\bf 39}, 216 (1960) [ Sov.\ Phys.\ JETP {\bf 12}, 155 (1961)].
\smallskip

\item{[14]}
B.I. Halperin, P.C. Hohenberg, and E.D. Siggia, Phys.\ Rev.\ B {\bf 13}, 
1299 (1976).
\smallskip

\item{[15]}
B. Yurke, A.N. Pargellis, T. Kovacs, and D.A. Huse, Phys.\ Rev.\ E {\bf 
47}, 1525 (1993).
\smallskip

\item{[16]}
M. Mondello and N. Goldenfeld, Phys.\ Rev.\ A {\bf 42}, 5865 (1990).
\smallskip

\item{[17]}
A. Bhattacharya and A. Chakravarty, unpublished.
\smallskip

\item{[18]}
K. Damle, T. Senthil, S.N. Majumdar, and S. Sachdev, EuroPhys.\ Lett. 
{\bf 36}, 7 (1996).

\end